\documentstyle[editedvolume]{crckapb}

\begin{opening}

\title{
SPACE DENSITIES FOR POWERFUL RADIO SOURCES
IN THE LIGHT OF UNIFICATION
}

\subtitle{}

\author{C.A. JACKSON}
\institute{Institute of Astronomy, University of Cambridge,\\
           Cambridge, CB3 0HA, UK}
\author{J.V. WALL}
\institute{Royal Greenwich Observatory, Madingley Road,\\
           Cambridge, CB3 0EZ, UK}
\end{opening}

\runningtitle{SPACE DENSITIES FROM UNIFIED MODELS}

\begin{document}


As radio survey frequency is raised the proportion of flat-spectrum sources
increases in bright flux-limited samples ({\it eg} Wall 1994, 
{\it Aust J Phys}
{\bf 47}, 625).  Differential source counts show a corresponding 
broadening of the
central maximum due to the increasing
proportion of flat-spectrum sources.  
Orr \& Browne (1982, {\it MNRAS} {\bf 200},
1067) modelled this change in shape of the source
count by proposing a unifying scheme 
which states that the core-dominated, flat-spectrum
radio sources are the steep-spectrum sources with their cores 
Doppler-boosted due to the alignment of the jets with the line of sight.

Investigation of the space densities of radio sources should proceed
with populations which are physically delineated; in the face of unified models,
the traditional division into `flat-spectrum' and `steep-spectrum' populations 
is incorrect. To this end we are undertaking a new space-density 
analysis to
explore the implications of unified-model schemes, including both
the radio-loud QSO -- FRII radio-galaxy paradigm and the BL\,Lac -- FRI 
radio-galaxy paradigm (see Urry and Padovani 1995, {\it PASP} {\bf 107}, 803). 
To test the formalism, our first
stage described here uses (1) complete samples and source-count
data over a wide frequency range and (2) optimizing techniques
to explore {\it parameterized} evolution and beaming models.

This initial analysis followed the scheme developed by Wall
{\it et. al} (1980, {\it MNRAS} {\bf 193}, 683).
Together with a 151-MHz source-count,
the 162 steep-spectrum sources in the 3CR sample (Laing {\it et. al}
1983, {\it MNRAS} {\bf 204}, 151)
were used to define the
epoch-dependent luminosity function of the
`parent' population.
The best-fit parameters were determined using the
{\it AMOEBA} downhill simplex method in multidimensions
(Press {\it et. al} 1992, {\it Numerical Recipes in Fortran} (CUP), 402),
evaluating $\chi^{2}$ between the observed and model source counts.
For evolution of the form $exp(M(1-t/t_0))$ the optimal
parameters ($\Omega = 1, h = 0.5$) are 
$M$=10.92, $z_{c}$=4.075 and transition powers between evolving and
non-evolving sources at log$_{10}(P_{1}) = 25.33$, log$_{10}(P_{2}) = 27.57$.
This demonstrates that modern data comprising complete redshifts for the 3CR
sources plus a deep source count {\it require} a redshift cut-off in the
space density for steep-spectrum sources.

These parameter values and a single spectral index of -0.75 were used
to estimate the 5~GHz count of steep-spectrum sources (Figure 1).
Inclusion of the flat-spectrum, 
beamed population at 5 GHz was achieved with two
additional parameters, the Lorentz factor $\gamma$ and the rest frame
core-to-extended flux ratio $R_{c}$.
The observed core-to-extended flux ratio $R_{obs}$ is given by
$R_{obs} = R_{c} ([\gamma(1 - \beta$cos$\theta)]^{-2+\alpha_{flat}} + 
[\gamma(1 + \beta
$cos$\theta)]^{-2+\alpha_{flat}})$ for a source comprising a pair of 
continuous relativistic
jets with bulk plasma velocity $\beta c$ whose ejection axis is 
aligned at a random angle $\theta$ ($\geq 0^{\circ}, \leq 90^{\circ}$) to
the line of sight. We adopted
$\alpha_{flat} = 0.0$, and took
a source as being `flat-spectrum' for small enough values of $\theta 
< \theta_c$ such
that $R_{obs} \geq 1.0$ and its
observed flux density $S_{enhanced} = R_{obs} . S_{\nu_{1}}$. 
For $\gamma = 10.0$ and $R_{c} = 0.02$ ($\theta_c = 8^\circ$), the
count of flat-spectrum sources 
summed with the steep-spectrum source count closely
follows the observed count (Figure~1).

\begin{figure}[h]
\vspace{2.2in}

\includegraphics{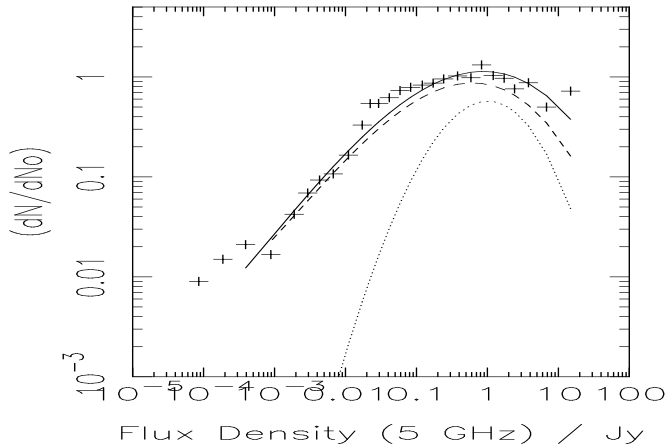}

\caption{Model and observed source counts at 5 GHz:}
\small $^{++++}$ observed source count,
\rule[0.05in]{0.04in}{0.01in}
\rule[0.05in]{0.04in}{0.01in}
\rule[0.05in]{0.04in}{0.01in}
\rule[0.05in]{0.04in}{0.01in} model count for steep-spectrum objects,
$\cdots$ model count for flat-spectrum objects,
\rule[0.05in]{0.25in}{0.01in} total model source count.

\end{figure}
\normalsize

This initial analysis demonstrates that (a) a diminution in the space density 
of `parent' sources at redshifts above 4 is required, and (b) the 
FRII -- radio-loud 
QSO unified scheme is consistent with the high-frequency count data for
reasonable beaming parameters.
 
\end{document}